\documentclass[twocolumn,preprintnumbers,endnote,nofootinbib,aps,prd,eqsecnum,floatfix]{revtex4}

\usepackage{footmisc}
\usepackage{amsmath,amssymb}
\usepackage{graphicx}

\usepackage{color}

\renewcommand{\d}{{\text{d}}}
\newcommand{\be}{\begin{eqnarray*}}
\newcommand{\ee}{\end{eqnarray*}}
\newcommand{\gl}[1]{(\ref{#1})}

\newcommand{\bee}{\begin{eqnarray}}
\newcommand{\eee}{\end{eqnarray}}
\newcommand{\beeq}{\begin{equation}}
\newcommand{\eeeq}{\end{equation}}
\newcommand{\cp}{${\cal{CP}}$}
\newcommand{\gev}{{\text{GeV}}}
\newcommand{\tev}{{\text{TeV}}}
\newcommand{\ifb}{{\text{fb}}^{-1}}
\newcommand{\fb}{{\text{fb}}}

\preprint{IPPP/12/86} \preprint{DCPT/12/172}

\begin{document}

\title{Polishing a shiny Higgs with matrix elements}

\begin{abstract}
  After the recent discovery of a Standard Model Higgs boson-like
  particle at the LHC, the question of its couplings to known and
  unknown matter is eminent. In this letter, we present a method that
  allows for an enhancement in $S/B$(irreducible) of the order of
  100\% in $pp\to (h\to \gamma\gamma) jj$ for a center of mass energy
  of 8 and 14 TeV. This is achieved by applying the matrix element
  method. We discuss the implications of detector resolution effects
  and various approximations of the involved event simulation and
  reconstruction.  The matrix element method provides a reliable,
  stable, and efficient handle to separate signal from background, and
  the gluon and weak boson fusion components involved in this
  process. Employing this method, a more precise Higgs boson coupling
  extraction can be obtained, and our results are of immediate
  relevance for current searches.
\end{abstract}

\author{Jeppe R. Andersen\vspace{0.2cm}}
\affiliation{Institute for Particle Physics Phenomenology, Department
  of Physics,\\Durham University, DH1 3LE, United Kingdom}

\author{Christoph Englert} 
\affiliation{Institute for Particle Physics Phenomenology, Department
  of Physics,\\Durham University, DH1 3LE, United Kingdom}

\author{Michael Spannowsky} 
\affiliation{Institute for Particle Physics Phenomenology, Department
  of Physics,\\Durham University, DH1 3LE, United Kingdom}

\maketitle


\section{Introduction}

The performance of the CERN LHC and the rapid collection of data by
both ATLAS and CMS will allow for detailed studies of the properties
of the recently discovered boson~\cite{:2012gk,:2012gu} in the coming
years. The first reported observations of the production cross section
and branching fractions are consistent~\cite{fits} with the
expectation from a Standard Model Higgs boson~\cite{orig} with the
observed mass $m_h\simeq 126~\gev$, but sizeable uncertainties caused
by the limited statistics are still present.

The projected amount of data collected this year will not only reduce
these uncertainties, but may also allow for the study of properties
beyond total rates. 

In many extensions of the Standard Model the existence of \cp-odd
bosons are predicted. Therefore, studying the \cp~properties of the
boson is of great importance to confirm if the mechanism of
electroweak symmetry breaking is minimal or not. A detailed study of
the kinematic distributions in the production of the boson in
association with two jets~\cite{Andersen:2008u,pertcol} will
eventually allow for the direct measurement of the \cp~of the
boson~\cite{hjjcoup,hjjcoup2}.

Equally important are direct measurements of the Higgs couplings to
all other particles of the Standard Model. In particular, the
loop-induced $h\gamma \gamma$ vertex is sensitive to new degrees of
freedom, where direct production might be beyond the energy reach of
the LHC. The number of observed signal events depends on the sum of
the production processes and the decay branching ratio:
\begin{equation}
  \sigma(h) \cdot BR(\gamma\gamma) 
  \sim (\sum_p g^2_p) \frac{g_{h\gamma\gamma}^2}{\sum_{\text{modes}} g_i^2} \,,
\end{equation}
assuming no interference between the different production channels $p$,
where $g$ denote the involved Higgs couplings. The sum in the
denominator runs over all kinematically accessible final states in the
decay. The precision in measuring any coupling of the Higgs boson
obviously benefits from separating the production mechanisms.

In this respect $pp\to hjj$ is important for two reasons. First we
observe an excess in $h\to \gamma\gamma$ in current analyses,
especially in the $2j$ category \cite{:2012gk}. Furthermore, an analysis of the $h\to
\tau\tau$ channel heavily relies on the $2j$ final state. For $pp\to
(h\to \tau\tau)jj$ we currently observe under-production \cite{:2012gu}. If both
these findings prevail and follow from new physics beyond the SM, a precise
investigation beyond simple ratios, which will also incorporate
modifications of the GF and WBF production modes, will majorly depend
on WBF/GF separation in this channel.

$pp\to (h\to \gamma \gamma)jj+X$ is roughly composed 50:50 by gluon
fusion (GF) and weak boson fusion (WBF), depending on, {\it{e.g.}},
the cut on the invariant dijet mass. In this context, it is important
to note that the quantum interference between the GF and WBF
components of $pp\to hjj+X$ production is completely
negligible~\cite{hjjint}, and hence it is possible to consider these
two contributions separately. Also, the theoretical uncertainty of the
GF contribution is much larger than it is for WBF production 
\cite{Campbell:2006xx,Figy:2003nv}, especially within the WBF-selection cuts.

In this letter, we construct a likelihood function which is based on
WBF and GF matrix elements which precisely serves this purpose to
separate the production mechanisms, {\it i.e.}  we apply the
\emph{matrix element method} \cite{mem} to the $pp\to (h\to
\gamma\gamma ) jj $ process on the fully-showered and hadronized final
state. We also generalize the GF~vs.~WBF vs.~background discrimination
in a realistic analysis by including the $pp\to \gamma \gamma +jj+X$
matrix elements. This constitutes the main irreducible backgrounds to
this search~\cite{:2012gk}. Employing this strategy allows to optimize
cut scenarios to enhance signal-over-background $S/B$. Given that the
irreducible $pp\to \gamma\gamma jj$ constitutes $\mathcal{O}(70)$\% of
the total background \cite{:2012je} an additional handle to reduce it
will greatly improve the sensitivity of the experimental analyses.

We organize this work as follows: First we discuss the matrix element
strategy for $pp\to (h\to \gamma\gamma) jj+X$ for the LHC
$\sqrt{s}=14~\tev$ in Sec.~\ref{sec:14tev}. This is the setup, for
which most precision Higgs results can be obtained from a large
luminosity sample ${\cal{L}}\gtrsim 300~\ifb/$experiment. We discuss
the systematic uncertainties of various approximations of the signal
event generation, especially for the GF contribution, and we also
include the impact of finite detector resolution.

Equipped with these insights, we discuss the implications of the
matrix element method for $pp\to (h\to \gamma\gamma)jj+X$ for the
8~TeV run and current results~\cite{:2012gk,:2012gu} in
Sec.~\ref{sec:8tev}. We conclude this work with a summary in
Sec.~\ref{sec:sum}.

\section{Higgs in association with two jets at 14 TeV}
\label{sec:14tev}

\subsection{The Matrix Element Method}
\label{sec:meme}
Let us introduce the observables that we are going to study in the
remainder of this paper. The GF/WBF discriminating likelihood is
defined
\begin{multline}
  \label{eq:signaln}
  {{\tilde{Q}}}_n(p_1^\gamma,p_2^\gamma,\{p^j_n\})\\
  =-\log\left[ \frac{\d {\text{LIPS}}(\gamma\gamma
      j^n) \, |{\cal{M}}^{\text{WBF}}(pp\to (h\to\gamma\gamma)j^n)|^2}
    {\d {\text{LIPS}}(\gamma\gamma
      j^n) \,  |{\cal{M}}^{\text{GF}}(pp\to (h\to\gamma\gamma)j^n)|^2}
  \right]\\ =
  -\log\left[ \frac{ |{\cal{M}}^{\text{WBF}}(pp\to (h\to\gamma\gamma)j^n)|^2}
    { |{\cal{M}}^{\text{GF}}(pp\to (h\to\gamma\gamma)j^n)|^2}
  \right] \,,
\end{multline}
where we denote the (partonic) jet multiplicity by $n$, $\d
{\text{LIPS}}$ is the differential phase space weight for the
particular kinematics (which is identical for all processes we
consider and hence drops out of the ratio), and $|{\cal{M}}|^2$
denotes the respective matrix elements, already include the parton
distribution functions (pdfs). We implement the CTEQ6l1 set
\cite{c6l1} in the ratio of Eq.~\gl{eq:signaln}.

Eq.~\gl{eq:signaln} provides a one-dimensional probability
distribution, which expresses statistical compatibility with either of
the two hypotheses in the best suitable way, by definition. By
including ${\tilde{Q}}_n$ to the event selection, we can optimize
${\tilde{Q}}_n\lessgtr\langle{\tilde{Q}}_n\rangle^{\text{WBF,GF}}$
depending on the purification requirement. The expectation values
$\langle . \rangle$ follow from MC simulations of signal and
background, similar to the construction of simple binned
log-likelihood ratio hypothesis tests~\cite{llhr}.

\begin{figure}[!t]
  \includegraphics[scale=0.7]{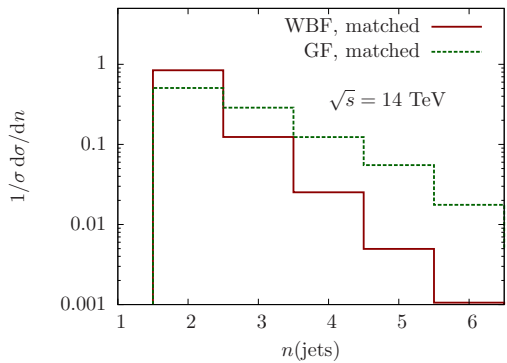}
  \caption{\label{fig:nj14} Exclusive number of jets distribution for
    the LHC running 14 TeV. The selection cuts are described in the
    text.}
  \vspace{0.2cm}
\end{figure}

We apply the effective top
approximation in the following for the GF contribution and the $h\to
\gamma\gamma$ decay via operators \cite{Kniehl:1995tn}
\begin{equation}
  \label{eq:efflag}
  {\cal{L}}\supset {\alpha_s \over 12\pi v } G^a_{\mu \nu}G^{a\,\mu\nu}h +
  {\alpha_{em} \over 2\pi v } F_{\mu \nu}F^{\mu\nu}h \left(
    e_t^2-7/4\right)\,.
\end{equation}
New degrees of freedom which alter the GF contribution and/or modify
the Higgs couplings to weak bosons can be included as a global factor
in the ratio of Eq.~\gl{eq:signaln}. A global factor merely shifts the
$\tilde Q$ by a constant factor, which is irrelevant for the
probabilistic discrimination of GF vs.~WBF. The model-specific details
enter the sampling of $\langle{\tilde{Q}}_n\rangle^{\text{WBF,GF}}$
and in the individual normalizations.

\begin{figure*}[!t]
  \includegraphics[scale=0.7]{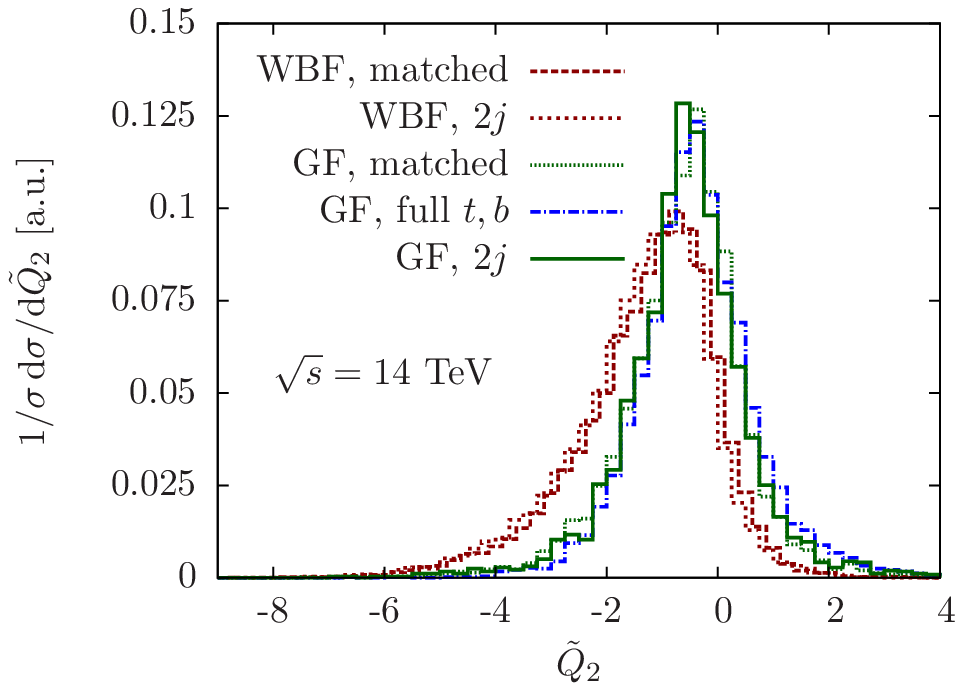}
  \hspace{1.5cm}
  \includegraphics[scale=0.7]{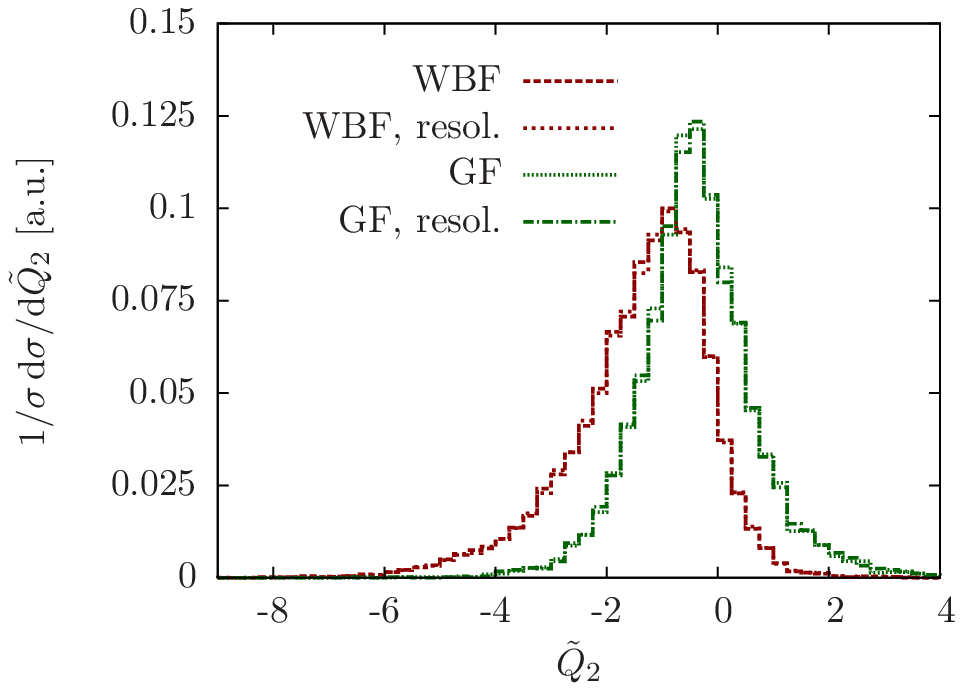} \\[0.2cm]
  \includegraphics[scale=0.7]{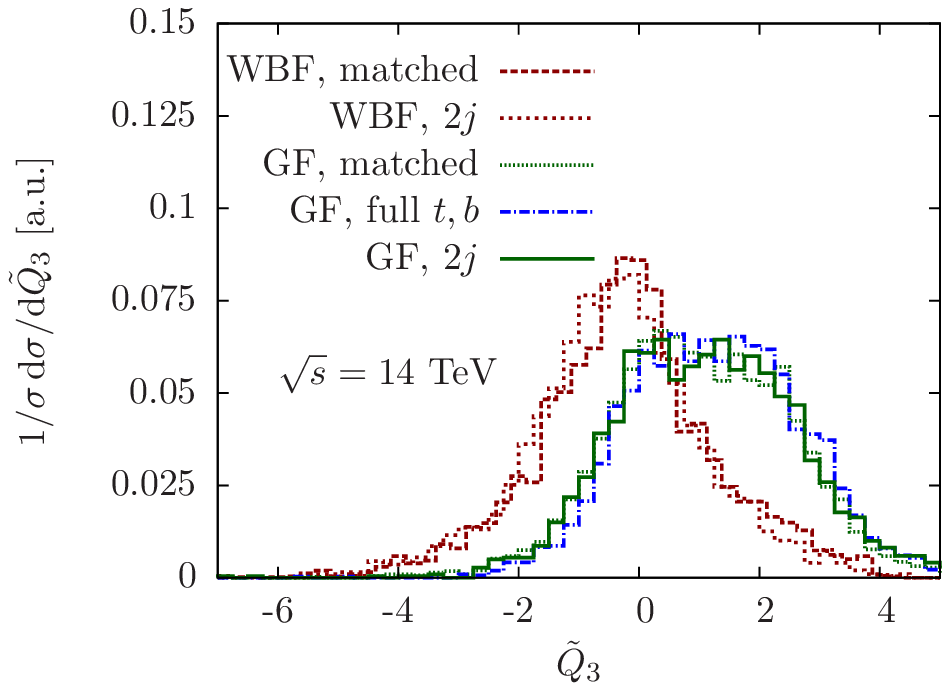}
  \hspace{1.5cm}
  \includegraphics[scale=0.7]{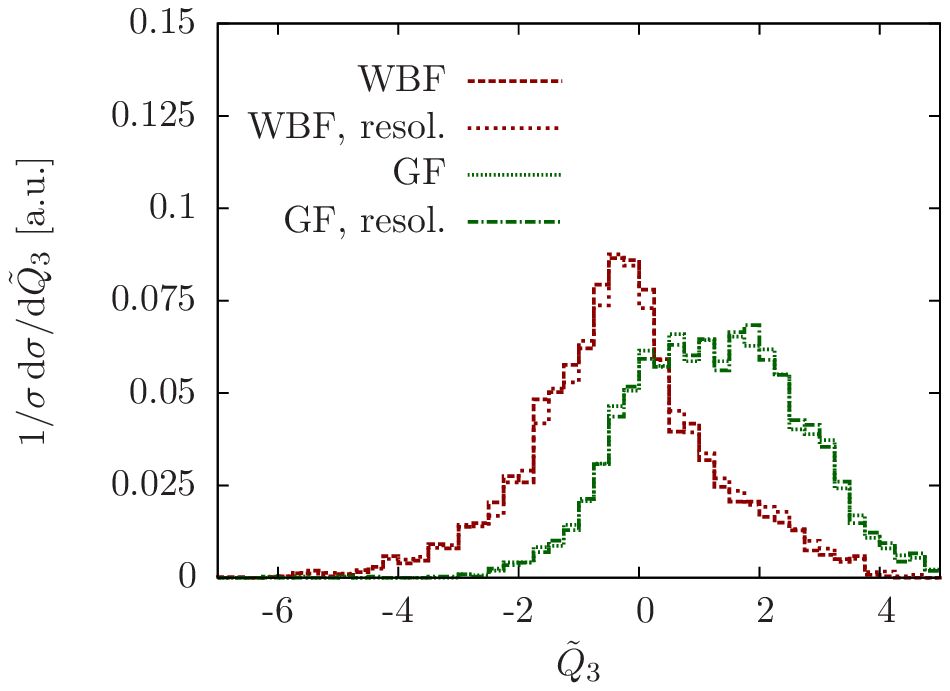}
  \caption{\label{fig:2jllh} 2-jet and 3-jet likelihoods $\tilde
    Q_{2,3}$ for the cuts as described in the text and
    $\sqrt{s}=14$~TeV. We show the influence of various event
    generation modes, where ``$2j$'' refers to generating $pp\to
    hjj\to \gamma\gamma jj$ events from 2 jet matrix elements+parton
    shower, ``matched'' refers to a matched $2j/3j$ sample, and ``full
    $t,b$'' stands for 2-jet events, including the full one loop mass
    dependence, interfaced to the parton shower. We also show the
    influence of detector and photon resolution
    effects.\vspace{0.3cm}}
\end{figure*}
\begin{figure*}[!t]
    \includegraphics[scale=0.7]{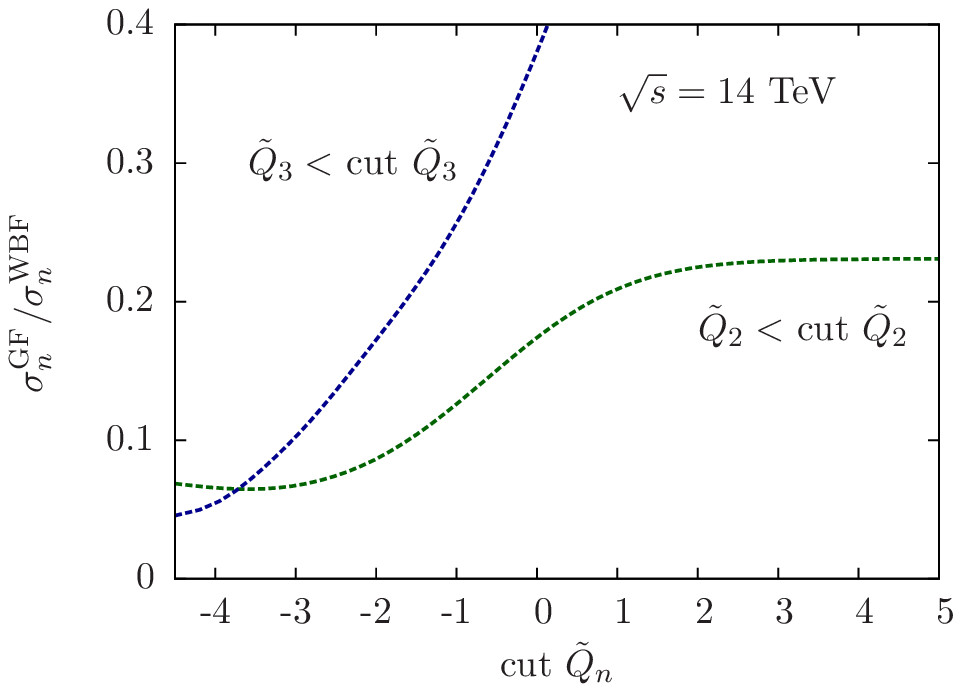}
    \hspace{1.5cm}
    \includegraphics[scale=0.7]{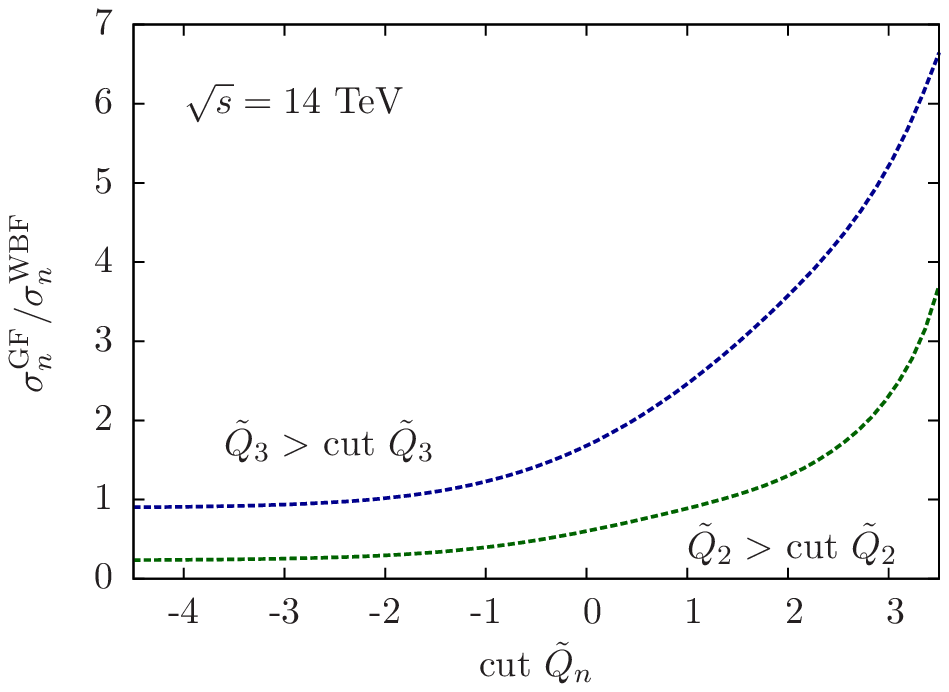}
    \caption{\label{fig:llhr14} Purification of GF vs.~WBF on the
      basis of the likelihood $\tilde Q_{2,3}$.}
\end{figure*}

The matrix elements that enter Eq.~\gl{eq:signaln} are functions of
parton-level kinematics and we have to define an algorithm which maps
the fully showered and hadronized jet final state to a suitable set of
(massless) kinematics, which also includes information about the
events initial state. We do this in the following way: We first
cluster jets with {\sc{FastJet}} \cite{fastjet} and reconstruct the
isolated photons according to the respective analysis requirements
(see further below). We count the number of jets passing the $p_{T,j}$
threshold in the events. We then re-distribute the transverse recoil
against unresolved radiation. The jets' momenta along the beam axis we
reconstruct from massless calorimeter cell entries at a given
pseudo-rapidity. We scale the energy of the resulting objects such
that $p^2=0$. From the sum of these objects, we get an overall energy
reconstructed boost of the considered particle system, which allows to
define two momentum fractions of initial state momenta. This way,
starting from an exclusive number of reconstructed jets $n_j$ and
photons $n_\gamma$, that comply with the analysis requirements we end
up with a set of parton level four momenta which we use for the
calculation of the ratios $\tilde Q_n$. This procedure is obviously
not limited to MC studies and can be incorporated by experiments
straightforwardly.

To discriminate signal from background we generalize
Eq.~\gl{eq:signaln} to the $S/B$, $S/\sqrt{B}$-improving likelihood
\begin{widetext}
  \begin{equation}  
    \label{eq:signalb}
    {{\tilde{Q}}}^b_n(p_1^\gamma,p_2^\gamma,\{p^j_n\})=
    -\log\left[ \frac{\left\{ |{\cal{M}}^{\text{WBF}}(pp\to
          (h\to\gamma\gamma)j^n)|^2 + 
          |{\cal{M}}^{\text{GF}}(pp\to (h\to\gamma\gamma)j^n)|^2\right\}}
      {|{\cal{M}}^{2\gamma}(pp\to \gamma\gamma j^n)|^2}  \right]\,.
  \end{equation}
\end{widetext}

For the numerical implementation of Eqs.~\gl{eq:signaln} and
\gl{eq:signalb} we rely on a combination of {\sc{MadGraph}} v4
\cite{madevent} and {\sc{Vbfnlo}} \cite{vbfnlo}.

\begin{figure}[!t]
    \includegraphics[scale=0.7]{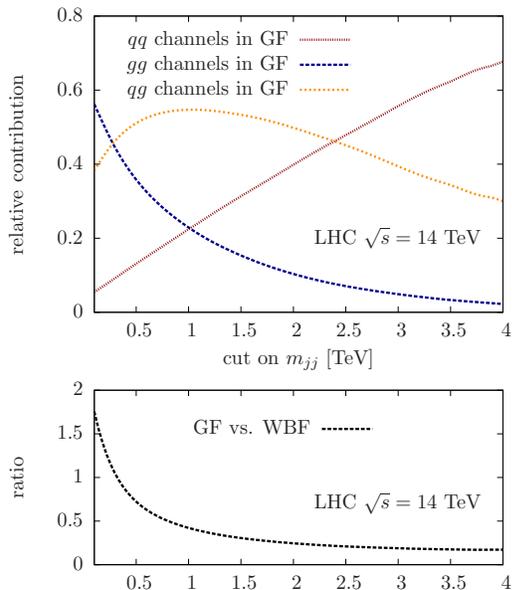}
    \caption{\label{fig:mjj14} $pp\to hjj$ production via gluon fusion
      (including full top and bottom contributions) broken down to the
      partonic channels and comparison of WBF (at NLO QCD) vs.~GF as a
      function of the cut on the invariant dijet mass. Results are
      obtained with {\sc{Vbfnlo}} \cite{vbfnlo}.}
\end{figure}

\subsection{Event generation and selection}
We generate two and three jet CKKW-matched \cite{ckkw} and un-matched
samples with {\sc{Sherpa}} \cite{sherpa}, which implements the
effective top approximation in the gluon fusion channels
\cite{Krauss:2001iv}. The events are generated with {\sc{Sherpa}}'s
default CT10~\cite{ct10} pdf set to avoid biasing the analysis of the
likelihood distributions.

It is known that the effective theory does not provide a valid
description of the phenomenology as soon as we are sensitive to
momentum transfers larger than the top mass, {\it e.g.} $p_{T,j}\geq
m_t$. Cross sections, on the other hand, are reproduced at the percent
level, which follows from smaller effective theory cross sections for
$p_{T,h}\lesssim m_t$ cancelling the excess with respect to the full
calculation for $p_{T,h}\gtrsim m_t$. For large momentum transfers
$p_{T,h}\sim 300~\gev$ the deviations become larger than
${\cal{O}}(30\%)$. To fully capture the dynamics in a two-dimensional
likelihood, which extends Eqs.~\gl{eq:signaln},~\gl{eq:signalb} by the
inclusion of the events energy scale would need to incorporate the
full mass dependence and a parametrization of potential new physics.

To test our effective theory approximation (note that we both simulate
and analyze the event within the effective approximation), we also
analyze a GF event sample obtained with {\sc{Vbfnlo}} \cite{vbfnlo},
which includes the full top and bottom contributions. We pass the
parton-level events to \hbox{\sc{Herwig++}}~\cite{Bahr:2008pv} for
showering and hadronization. This allows us to asses the bias of
analyzing the events with effective theory matrix elements.

Finite detector resolution effects give rise to an additional
systematic uncertainty when we want to analyze the discriminating
power of the matrix element method. To assess their impact on
Eq.~\gl{eq:signaln} and \gl{eq:signalb}, we model jet momentum
uncertainties according to Ref.~\cite{atlastdr}:
\begin{equation}
    \label{eq:resolj}
    {\Delta E\over E} = {5.2\over E} \oplus
    {0.16\over
      \sqrt{E}} \oplus 0.033\,.
\end{equation}
For the photons we have performed a comparison of the invariant
diphoton mass against the experimental results of
\cite{:2012gk,:2012gu} and find that a resolution uncertainty
parametrized by 3\% describes the experimental situation in
sufficiently well for our purposes.

Details of the 14 TeV cut set-up are currently unknown; they will
depend on the new pile-up and underlying event conditions as well as
on the potential excess in the $pp\to (h\to \gamma\gamma) jj+X$ signal
cross sections. For the remainder of this section, we therefore adopt
a cut setup which is based on phenomenological analyses (see
{\it{e.g.}}  Refs.~\cite{hjjcoup,Hankele:2006ma,kinem}). In
Sec.~\gl{sec:8tev} we employ the current 8 TeV selection of
ATLAS~\cite{:2012gk}.

We reconstruct $k_T$ jets \cite{soper} with $D=0.8$ and threshold
$p_{T,j}\geq 30~\gev$ in the pseudo-rapidity range $|\eta_j|<4.5$, and
require $n_j \geq 2$. The exactly two isolated photons are required to
be central in the electromagnetic calorimeter $|\eta_\gamma|<2.5$ with
$p_{T,\gamma}\geq 30~\gev$ (we define a photon to be isolated if the
electromagnetic and hadronic activity in the cone with size $R=0.3$ is
less then $10\%$ of the $E_{T}$ of the photon candidate). The photons
have to reconstruct the Higgs mass $m_h=126~\gev$ within $115~\gev\leq
(p_{\gamma,1}+p_{\gamma,2})^2 \leq 135~\gev$. On top of these generic
cuts we impose a typical WBF selection: The two jets, leading in $p_T$
are required to have a large invariant mass $m_{j_1j_2}\geq 600~\gev$
and have to fall into opposite detector hemispheres $y_{j_1}\times
y_{j_2}\leq 0$.

\begin{figure*}[!t]
 \includegraphics[scale=0.7]{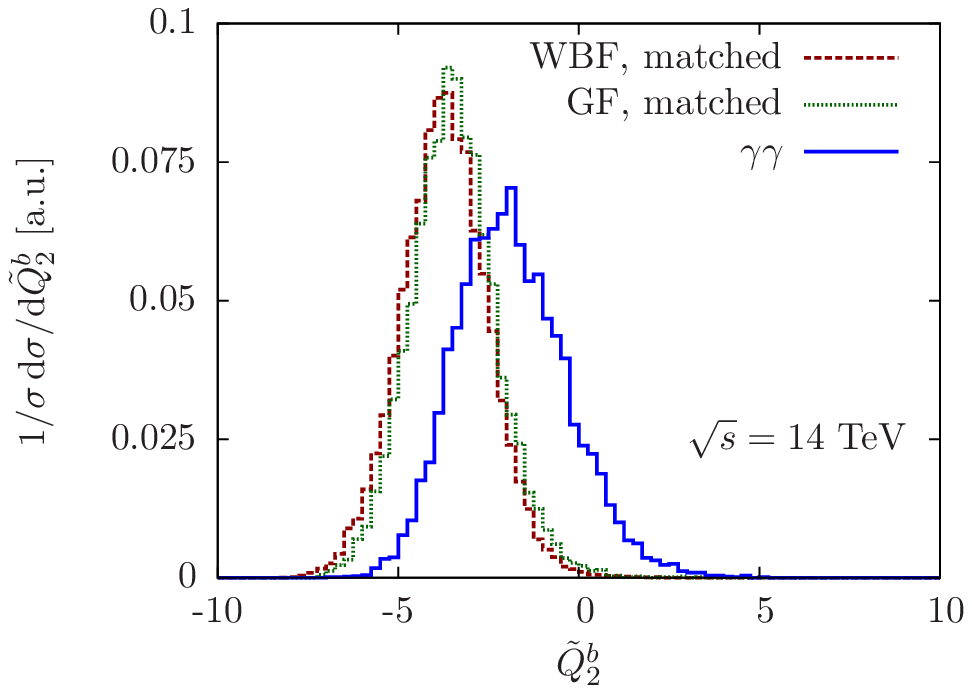}
 \hspace{1.5cm}
 \includegraphics[scale=0.7]{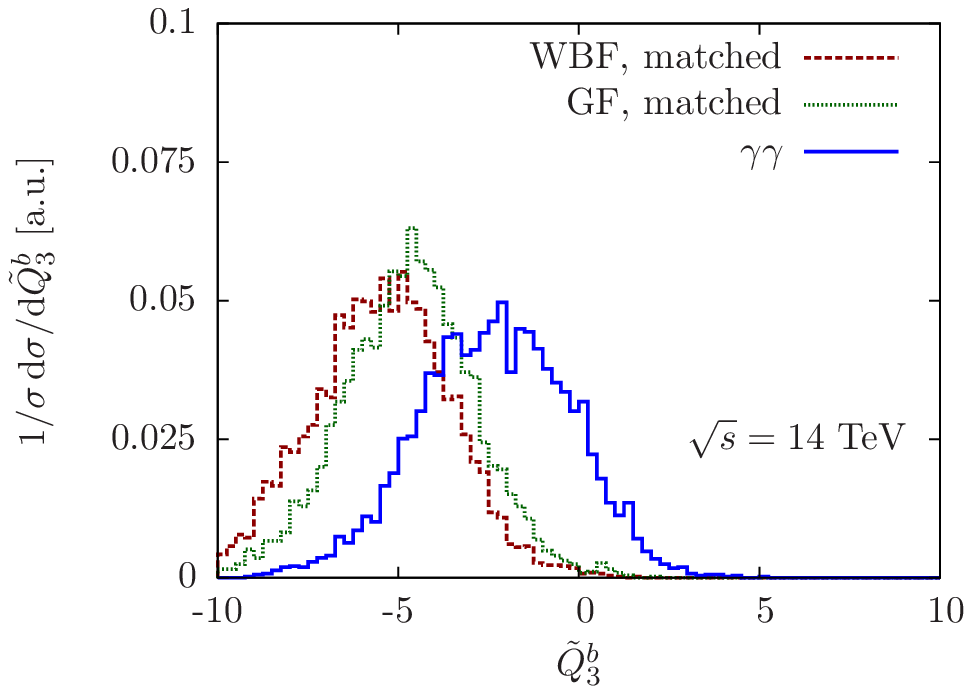} \\[0.2cm]
 \includegraphics[scale=0.7]{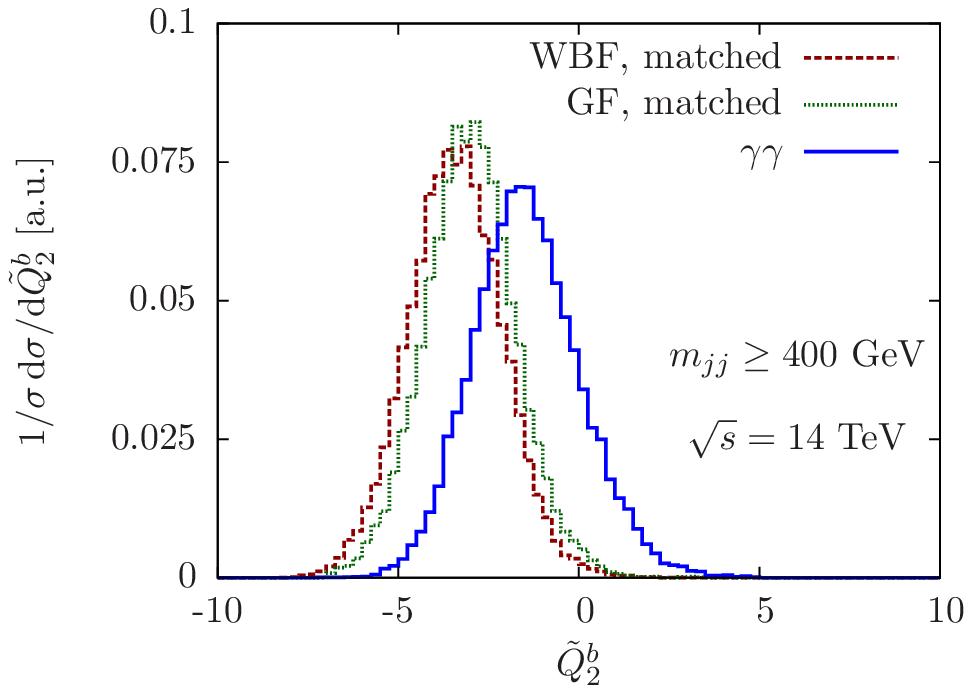}
 \hspace{1.5cm}
 \includegraphics[scale=0.7]{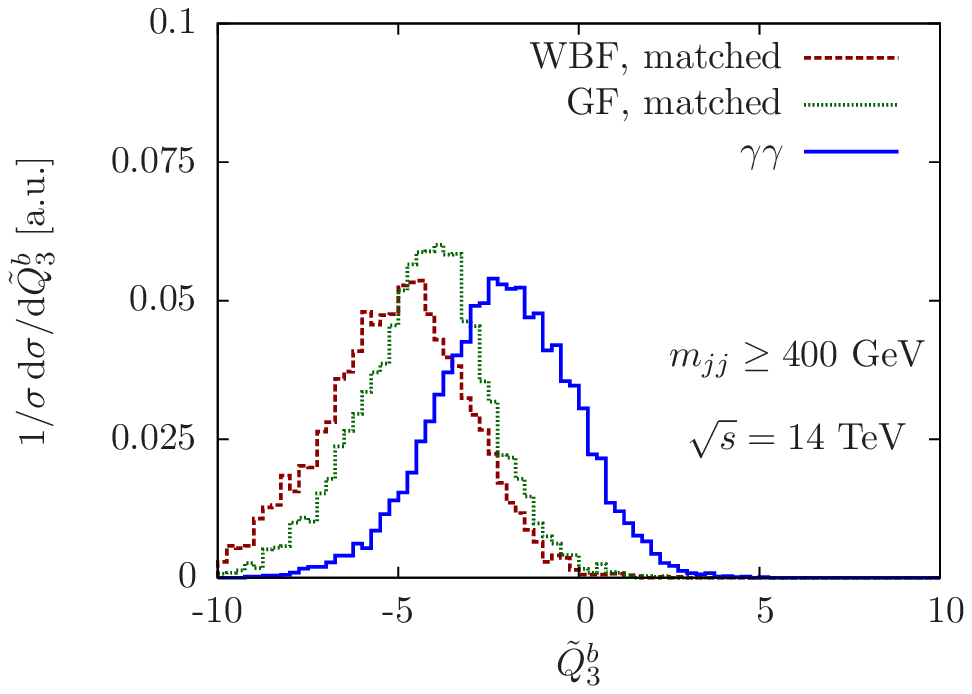} 
 \caption{\label{fig:14bkg} 2-jet and 3-jet likelihoods $\tilde
   Q^b_{2,3}$ for the cuts as described in the text and
   $\sqrt{s}=14$~TeV. Detector resolution effects are included. The
   lower panel gives results for a loosened cut set with $m_{jj}\geq
   400\gev$.}
\end{figure*}

\subsection{Performance of $\tilde Q_n$}
\label{sec:perf14}
For these 14 TeV selection criteria we show the normalized
distribution of the exclusive number of jets for 14 TeV in
Fig.~\ref{fig:nj14} (for a detailed analysis of this observable see
Ref.~\cite{higgsscale}). The cross sections are
$\sigma^{\text{GF}}\simeq 0.61~\fb$ and $\sigma^{\text{WBF}}\simeq
1.58~\fb$. From this we see that we can limit our analysis to $\tilde
Q_n,\tilde Q^b_n$ for $n=2,3$. Higher jet multiplicities contribute at
a level which is challenged by the theoretical uncertainties of the
inclusive $p\to hjj+X$ cross sections
\cite{Campbell:2006xx,Figy:2003nv}. The extension to $n>3$ is
technically straightforward.

We show our results in Fig.~\ref{fig:2jllh}. An immediate first
observation is that neither the definition of the likelihoods nor the
impact of either detector resolution or details of the event
simulation have a significant impact on the discriminating power of
$\tilde{Q}_{n=2,3}$. Keeping in mind that we already look at WBF-like
events and have a reconstructed Higgs boson, a cut on $\tilde Q_2$
will significantly enhance the WBF contribution over GF
production. This is even more the case for $\tilde Q_3$. This can be
understood along the following lines. Additional jet radiation in the
WBF component\footnote{For a detailed discussion of this contribution
  beyond LO see Ref.~\cite{Figy:2007kv}.} is essentially QCD
bremsstrahlung off the leading jets since there is no color exchange
between the quark legs. This is entirely different for the GF
contribution, which tends to populate the large available phase space
in the central region with QCD activity \cite{Dokshitzer:1991he}. Not
only the presence of this radiation~\cite{Cox:2010ug}, but, more
importantly, the information that is encrypted in the differential
energy-momentum flow associated with it \cite{hjjcoup2} provides an
elaborate handle to separate WBF from GF. This is most efficiently
reflected in $\tilde{Q}_{n\geq 3}$, of which only $n=3$ is
statistically relevant\footnote{The event shape observables discussed
  in Ref.~\cite{hjjcoup2} also capture sensitivity from relatively
  soft radiation which, by definition of $\tilde{Q}_n$ as jet based
  observables is not considered.}.

The result of GF/WBF purification is depicted in
Fig.~\ref{fig:llhr14}. Going to large values $|\tilde Q_{2,3}| \gg 1$,
we recover a high level of purification within the limits of the
selection criteria as expected,
$\sigma^{\text{GF}}/\sigma^{\text{WBF}} \lesssim 0.1$,
$\sigma^{\text{GF}}/\sigma^{\text{WBF}} \gtrsim 3$, respectively.
Comparing to Figs.~\ref{fig:2jllh}, this comes at the price of
increasingly small signal cross sections. A more realistic analysis
(that also includes a more realistic description of the detector
environment, pile-up, etc.)  needs to optimize the signal rate with
respect to purification. Applying multiple cuts on $\tilde Q_n$ for an
identically chosen basic cut set-up, on the other hand, can be used to
define distinct signal regions which allow to reconstruct the WBF/GF
content upon correlating the two measurements.

\begin{figure*}[!t]
  \includegraphics[scale=0.7]{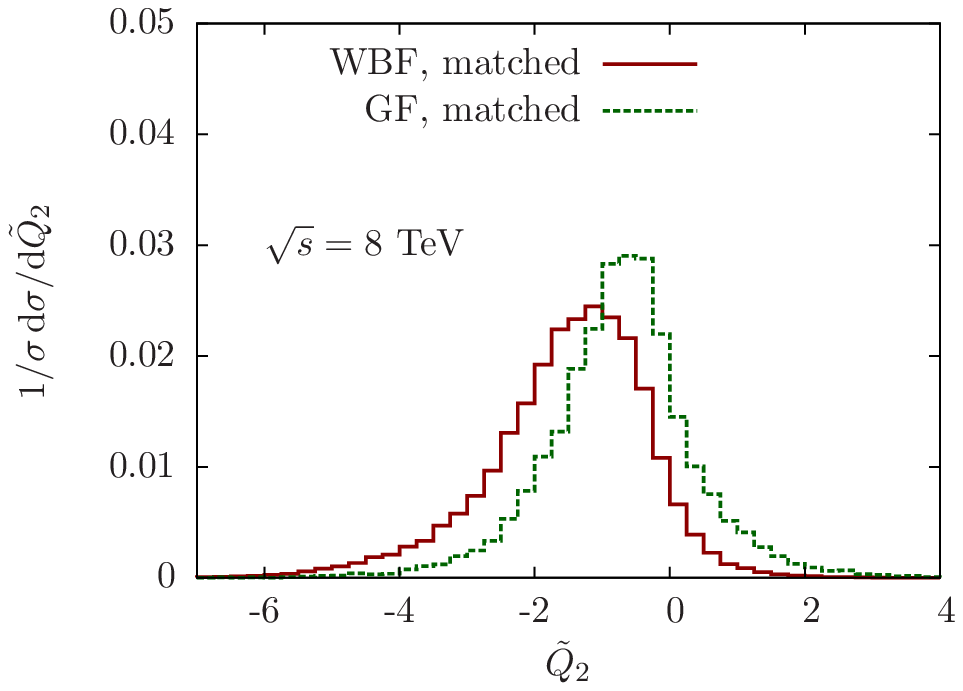}
  \hspace{1.5cm}
  \includegraphics[scale=0.7]{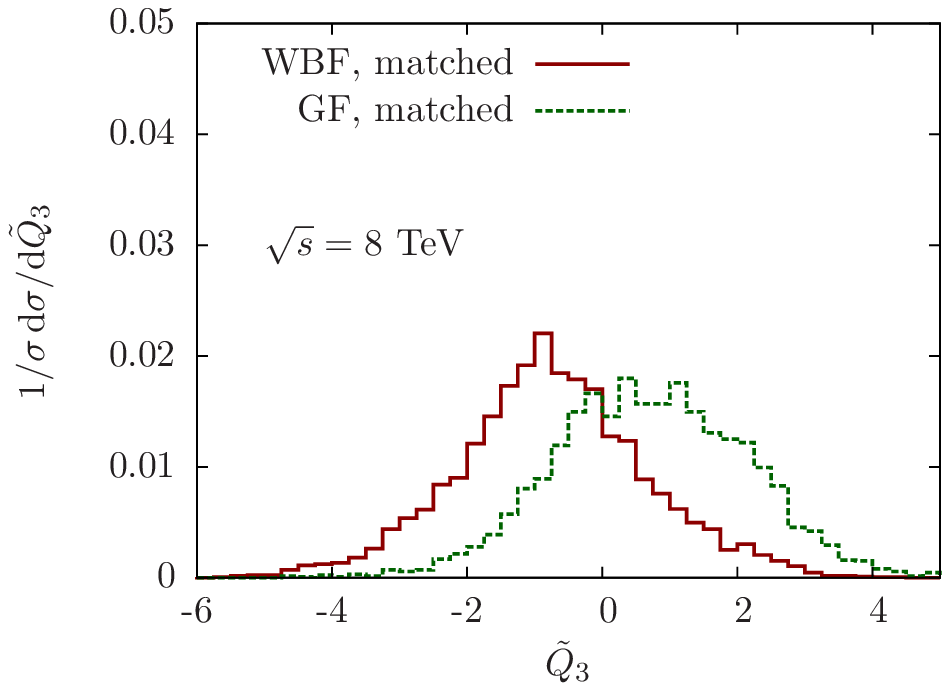}\\[0.2cm]
  \includegraphics[scale=0.7]{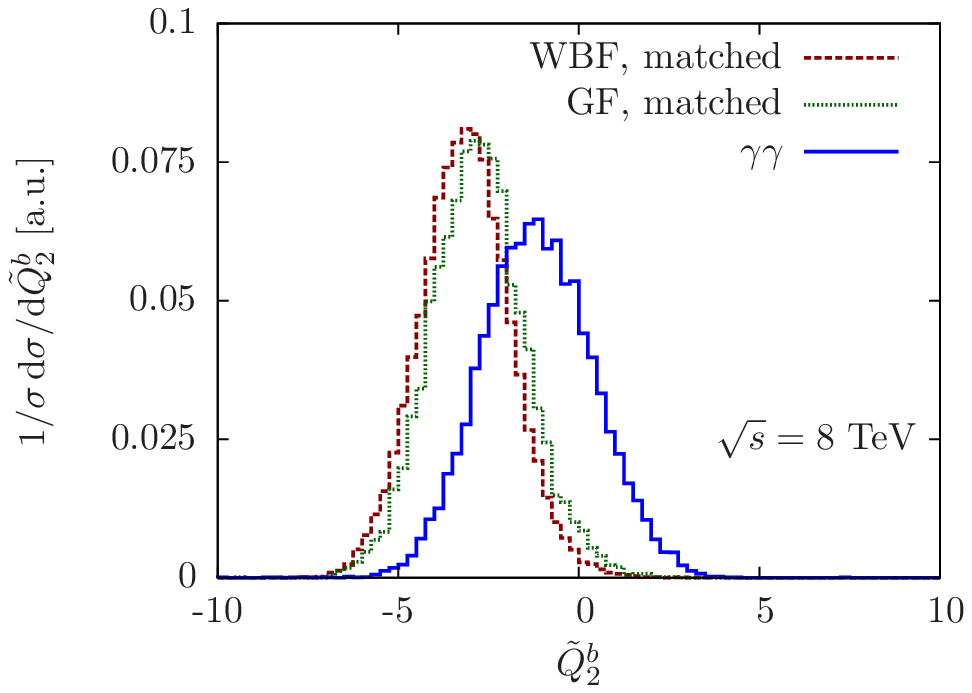}
  \hspace{1.5cm}
  \includegraphics[scale=0.7]{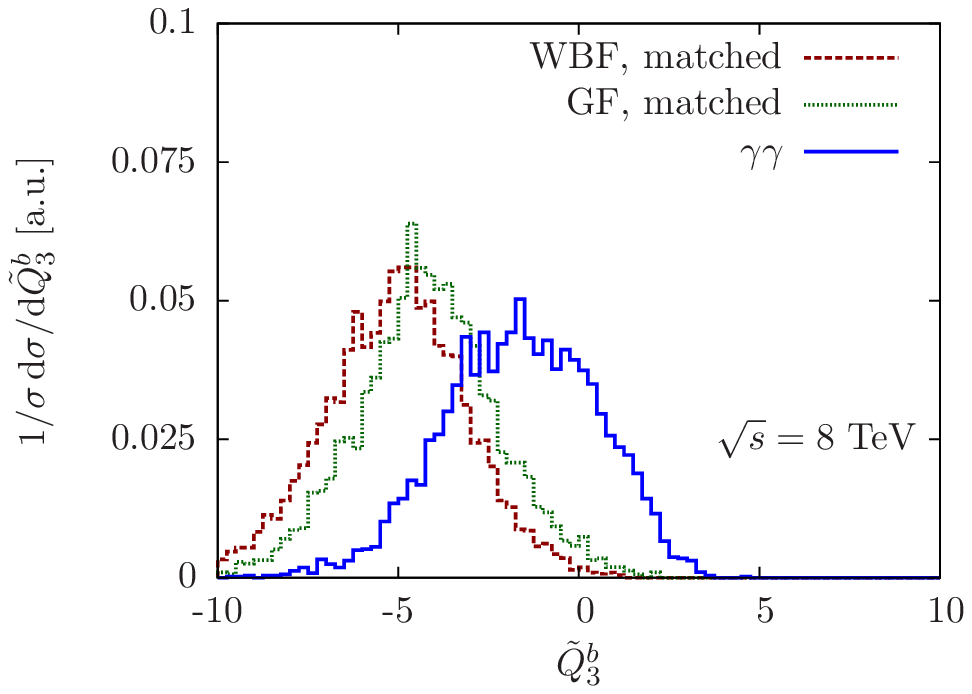}
  \caption{\label{fig8tev_result} The matrix element observables
    $\tilde{Q}_2,\tilde{Q}_3,\tilde{Q}_2^b,$ and $\tilde{Q}^b_3$ for 8
    TeV, employing the Higgs search' 2-jet category cuts of
    Ref.~\cite{:2012gk}.}
\end{figure*}

This finding needs to be contrasted to the standard paradigm of WBF/GF
separation via a larger $m_{jj}$ cut. In Fig.~\ref{fig:mjj14}, we show
the relative fraction of full one-loop GF vs.~NLO QCD WBF for
$\sqrt{s}=14~\tev$ as a function of the imposed $m_{jj}$ selection. We
see that the emerging quark-induced processes at large $x$ for a stiff
$m_{jj}$ cut saturate the relative fraction at about $15\%$. The
resulting event topology is WBF-like at small total cross
sections. Central jet vetos can further decrease the relative
contribution, but challenges the perturbative description
\cite{pertcol}.

\bigskip

We now turn to signal vs.~background discrimination. In
Fig.~\ref{fig:14bkg} we show the distributions for $\tilde Q_{2,3}^b$
for our $\sqrt{s}=14~\tev$ selection. Obviously, by cutting on $\tilde
Q_{2,3}^b$, we can purify the signal+background sample without loosing
too much of the signal count. This way the signal compared to the
irreducible background can be increased by $>100\%$. It is worth
noting that this will affect the GF and WBF contribution almost
identically (this is the reason why we choose to plot WBF and GF as
separate samples in Fig.~\ref{fig:14bkg}).

In order to gain a better handle on the GF distributions we can relax
the $m_{jj}$ cut such that the event is not forced into a WBF-type
topology. In Fig.~\ref{fig:mjj14}, we also show distributions for
$m_{jj}\geq 400\gev$. As expected, the difference in the WBF/GF
samples is now more pronounced in $\tilde Q^b_n$. The irreducible
$\gamma\gamma$ background grows by approximately a factor of two while
the gluon fusion and weak boson fusion cross sections are $0.95~\fb$
and $2.1~\fb$. This is an increase by factors $1.6$ and $1.3$
respectively, compared to the $m_{jj}\geq 600~\gev$ selection. We find
similar WBF/GF separation properties as in Fig.~\ref{fig:llhr14}. Note
that the irreducible background grows disproportionally, and a signal
vs.~background enhancement will vitally depend on a good $S/B$
discriminator, which is exactly provided by $\tilde Q_n^b$.

\begin{figure}[!t]
    \includegraphics[scale=0.7]{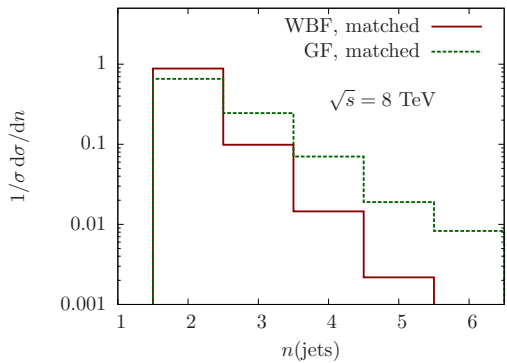}
    \caption{\label{fig:nj8} Exclusive number of jets distribution for
    LHC 8 TeV.}
\end{figure}

\section{Application to Higgs in association with two jets at 8 TeV}
\label{sec:8tev}

We can straightforwardly adopt the strategy of Sec.~\ref{sec:meme} to
the current 8 TeV setup. The ATLAS selection for the two jet category
of the $h\to \gamma \gamma$ search is as follows \cite{:2012gk}: We
cluster anti-$k_T$ jets \cite{antikt} with {\sc{FastJet}}
\cite{fastjet} for $D=0.4$ and select at least two jets with
$p_{T,j}\geq 25~\gev$ and $p_{T,j}\geq 30$ in the more forward region
$2.5\leq |\eta_j|\leq 4.5$. The hardest jets are required to have a
rapidity gap $|\Delta \eta_{jj}|\geq 2.8$ and the dijet system has to
recoil against the diphoton system in the transverse plane
$\Delta\phi(jj,\gamma\gamma)\geq 2.6$. Again as in
Sec.~\ref{sec:14tev} we require a Higgs mass reconstruction within 20
GeV interval centered around $m_h=126~\gev$.

The exclusive number of jets for this selection is again shown in
Fig.~\ref{fig:nj8}; and we find agreement of our analysis with the
experiment's quoted number of 3 expected events in
$4.7~\ifb$. Obviously, there is again no need to go beyond $n=3$.

\bigskip

Finally we again analyze the potential $S/B$ improvement (where $B$
refers to the irreducible background for our purposes), which is the
key limiting factor when dealing with the small event rates for the 8
TeV run. Fig.~\ref{fig8tev_result} shows a similar behavior as
Fig.~\ref{fig:14bkg}, we infer that we can at least gain a factor of
$100\%$ in $S/B$ without cutting into the signal count in the
currently applied selection. All remarks of the 14 TeV results
generalize to the lower energy of 8 TeV, and again the GF and WBF
signals rates are affected identically by selecting events according
to ${\tilde Q}^b_{2,3}$.

\section{Summary and Conclusions}
\label{sec:sum}
In this paper, we have applied the matrix element method to
$pp\to(h\to \gamma \gamma)jj$ production and investigated the
prospects to separate the GF and WBF contributions. This is of utmost
importance for \cp~analyses of the newly discovered particle, as well
as for the measurement of its couplings to known matter. The same
method can be applied to other decay modes of the Higgs boson, {\it
  e.g.} $h \to \tau \tau$.

We find that the matrix element method provides an excellent
discriminator, which is stable against finite detector resolution
effects and event generation systematics. We have extended WBF/GF
separation to signal vs.~(irreducible) background discrimination an
find promising results. WBF/GF separation and enhanced signal
vs.~background discrimination can be achieved simultaneously. Our
results are directly relevant for the current 2-jet category of $h\to
\gamma\gamma$ analysis, improving the sensitivity to the signal over
the irreducible $\gamma \gamma $ background by $\gtrsim 100\%$, thus
leading to a potential sensitivity increase by 35\% when reducible
backgrounds are considered unaltered.

Already in the presently available data sets using ${{\tilde{Q}}}^b_n$
the event selection cuts can be relaxed to boost the signal event
count, and thus significance, without affecting $S/B$.

Selection strategies which are based on $\tilde Q_n^{(n)}$ in a
experimentally realistic analysis will allow to measure Higgs
properties to better precision than currently foreseen. The
implementation of the method will become publicly available to the
experiments.

\medskip

{\it Acknowledgements:} We thank Frank Krauss for helpful and
entertaining conversations. CE acknowledges funding by the Durham
International Junior Research Fellowship scheme.

\end{document}